\documentstyle[aps,preprint]{revtex}
\draft

\begin{document}

\newcommand{\uprule}{\end{multicols}
\noindent \vrule width3.375in height.2pt depth.2pt 
\vrule height.5em depth.2pt \hfill \widetext }
\newcommand{\downrule}{\indent \hfill \vrule depth.5em height0pt 
\vrule width3.375in height.2pt depth.2pt 
\begin{multicols}{2} \narrowtext}	
	
\include{psfig}

\draft \title{\bf Local quantum coherence and superfluidity}

\author{F. de Pasquale $^{a,b}$ and S.M. Giampaolo $^{a}$}

\address{\mbox{$^a$ 
Department of Physics University of Rome 
"La Sapienza", P-le A.Moro 5, 00185 Rome Italy }} 
\address{\mbox{$^b$ INFM section of Rome}}

\date{Mar. 14 2001} \maketitle

\begin{abstract}
We consider a model of bosons on a regular lattice with a kinetic energy due to
hopping among sites and a potential energy due to strong on site interaction.
A superfluid phase is expected when the ground state of the local energy 
is doubly degenerate. We consider a new scheme of simmetry breaking associated
to the superfluid phase in which the order parameter is the statistical average
of the quantum coherence operator associated to 
the superposition of the degenerate local ground states. 
In the strong coupling limit a systematic expansion of the free energy can be 
performed in terms of the hopping amplitude at constant order parameter. 
Within such an expansion we obtain a self-consistent equation for the order 
parameter. The first order approximation gives, in the case of degeneracy 
between single occupied and empty state, the same result of the standard mean 
field approximation for the ``hard core bosons''. This new approach to the 
superfluid phase is shown to have a natural application to the implementation 
of quantum computation on a superfluid. 
\end{abstract}
The achievement of Bose Einstein Condensation experiments 
\cite{dalfovo,friebel,hamann},
and the interest to the quantum computing lead naturally to the proposal of 
using superfluid macroscopic states as qubits\cite{yushi}. On the other hand 
trapping of bosons in optical lattices,\cite{jaksch} opens the possibility of
new gating procedures. It is therefore of interest to 
improve the theory of bosonic particles with strong repulsive interaction and 
hopping among sites of a regular lattice and to investigate local excitations 
as possible qubits.

It is well known that at zero temperature the 
transition from a superfluid to a Mott insulator phase is expected
at a critical value of the ratio between the interaction strength and hopping 
amplitude except for integer values of the ratio between the chemical 
potential and the interaction strength $\mu=Un^*$ \cite{fisher,sheshandri}. 
These values correspond in the atomic limit, i.e. for vanishing hopping 
amplitude, to a degeneracy of the ground state which corresponds to the same 
statistical weight of states involving $n^*$ and $n^*+1$ particles. The 
associated energy can be assumed to vanish for a suitable choice of the 
zero energy level. A strong coupling expansion around the atomic limit has 
been proposed in \cite{sheshandri}. This approach depends however on a 
``factorization'' assumption. We show that an alternative approach can be 
introduced which avoids any factorization, and gives, to the first order in 
the hopping amplitude, the classical mean field result. The key point is a new
definition of the superfluid phase in which the order parameter is associated 
to the statistical average of suitably defined local quantum coherence 
operator.
 
We define the coherence of a system with respect to a particular 
state as the probability of finding the system, or part of it, in that state 
in the ensemble of all possible states. If the system is defined on a lattice, 
as in our case, and the ensemble is that of the statistical equilibrium, it is
possible to define a local quantum coherence probability {\mbox {( LQCP )}} 
as the average of the projection operator $\pi_i=|A_i><A_i|$ associated to the 
local state. 
\begin{equation}
\rho(\pi_i)=\frac{tr \left( e^{-\beta H} \pi_i \right)}
{tr \left(e^{-\beta H} \right)}
\end{equation}
H is the energy operator in the grand canonical ensemble. It is the sum
of a local repulsive term $H_0=\frac{U}{2} \sum_i n_i(n_i-1) - \mu\sum_i n_i$ 
( $U>0$ ) and a kinetic one $H_1=-J\sum_{ik}\epsilon_{ik}b^+_ib_k$ where 
$J>0$ and $\epsilon_{ij}$ is equal to $0$ if site i is not next neighbour of 
site j and 1 otherwise.
We consider here $|A_i>$ as a superposition of the two degenerate groundstates
of the local energy operator. In the subspace of two degenerate groundstates it
is useful to define a Pauli vector operator whose components are 
\begin{eqnarray}
\label{sigma1}
\sigma_{i1} & = &  \left( |n^*+1><n^*| +|n^*><n^*+1| \right) \nonumber \\
\sigma_{i2}& = & \frac{1}{i}\left( |n^*+1><n^*|-|n^*><n^*+1| \right) 
\nonumber \\
\sigma_{i3}& = & \left(|n^*+1><n^*+1|-|n^*><n^*| \right) 
\end{eqnarray}
The identity restricted to this subspace is defined as
\begin{equation}
\sigma_{i0}  =   \left( |n^*+1><n^*+1| + |n^*><n^*| \right) \nonumber \\
\end{equation}
Analogously to the spin case\cite{FEM} the LQC operator can be expressed in 
terms of the restricted identity and the scalar product of a Pauli operator 
with the unitary vector ( $ \nu_i$ ).  
\begin{equation}
\pi_i= \frac{1}{2}\sigma_{i0}+\sigma_i \cdot \nu_i
\end{equation} 
We show that superfluidity can be associated to a
non trivial LQCP ( $\rho(\pi_i)\neq \frac{1}{2}$ ). Differently from the 
classical approach the gauge symmetry breaking is achieved introducing a field
defined in the degenerate groundstate subspace in the 1-direction. It is then 
natural to assume this direction as polar axis and thus $\pi_i$ becomes 
\begin{equation}
\label{pigreco}
\pi_i=\frac{1}{2} \left[ \sigma_{i0} + \cos\theta_i \sigma_{i1} + 
\sin \theta_i \left( e^{i\phi_i} \sigma_i^- + e^{-i\phi_i} \sigma_i^+ 
\right) \right]
\end{equation}
Raising and lowering operators $\sigma_i^{\pm}=\sigma_{i2}\pm i \sigma_{i3} $ 
appear in (\ref{pigreco}). Spontaneous gauge symmetry 
breaking is studied in terms of the generating function associated to the 
1-direction in the degenerate groundstates subspace.
\begin{equation}
\label{defG}
G= tr \left[ e^{-\beta H}exp\left( \sum_i \lambda_i \sigma_{1i} 
\right)  \right]
\end{equation}
The local order parameter m is defined as $m_i=\frac{\partial}{\partial 
\lambda_i}\ln G$. Symmetry breakdown occurs if the order parameter does not 
vanish 
in the limit of $\lambda_i \rightarrow \lambda \rightarrow 0 $. The next step 
is to introduce the free energy as Legendre transform of $\ln G$
\begin{equation}
A=\ln G-\sum_i \lambda_i m_i
\end{equation}
The free energy A depends on $m_i$ instead of the symmetry breaking amplitude 
$\lambda_i$, and $\lambda$ becomes a function of $J$ determined by the 
condition
\begin{equation}
\label{lambda}
\lambda_i(J)=- \frac{\partial A }{\partial m_i}
\end{equation}
Note that now the order parameter $m_i$ does not depend on the hopping 
amplitude $J$ and thus can be fixed from the relation with $\lambda$ valid at
$J=0$   
\begin{equation}
m_{i}=\frac{ tr \left\{ e^{-\beta H}exp\left[ \sum_i \lambda_i(0)
\sigma_{1i} \right] \sigma_{1i} \right\}  }
{ tr \left\{ e^{-\beta H}exp\left[ \sum_i \lambda_i(0)
\sigma_{1i}  \right] \right\}}=<\sigma_{1i}>
\end{equation}
Following the work of \cite{georges} we introduce an expansion in the reduced 
hopping amplitude $\beta J$. The difficulty is 
due to the presence, unlike the Ising case, of an atomic Hamiltonian which 
does not commute with the hopping one. Standard formalism of field theory is 
not suitable for the expansion of the generating function of (\ref{defG}). 
It is worth to note that if the trace is performed 
on the eigenstates of the atomic Hamiltonian, only the degenerate groundstates
are affected by the symmetry breaking operator. The expansion in power of 
$\beta J$ around the value $\beta J=0$ of free energy A can be expressed as 
follows
\begin{equation}
A= A_0 + A_1
\end{equation}
Where $A_0$ is defined as  
\begin{equation}
\label{A0}
A_0 = \ln tr \left\{ e^{-\beta H_0}exp\left[ \sum_i \lambda_i(0)\left( 
\sigma_{1i} - m_i\right) \right] \right\}
\end{equation}
It is immediately shown that 
\begin{equation}
A_0= \sum_i \ln \left\{ 2\left[\cosh\left(\lambda_i(0) 
\right) - \gamma(\beta U) \right]\right\} -m_i \lambda_i(0) 
\end{equation}
where $\gamma(\beta U)=1 - \frac{1}{2}\sum_k e^{-\beta E_k}$. 
It is important to note that in the strong coupling limit ( $\beta U \gg 1$ ) 
$\gamma(\beta U)$ vanishes. In this limit the result valid for the hard core 
boson model is recovered. $A_1$ can be calculated as power expansion in the 
reduced hopping amplitude $\beta J$
\begin{equation}
A_1= A' (\beta J)+ A'' \frac{(\beta J)^2}{2} + \cdots 
\end{equation}
where 
\begin{equation}
A' = \sum_{ik}\epsilon_{ij} m_i m_k (n^*+1)
\end{equation}
while $A''$ has been calculated up to the first order in $\frac{1}{\beta U}$ 
\begin{equation}
A'' = \frac{(n^*+1)^2}{\beta U}\sum_{ikl}\epsilon_{ik}\epsilon_{il} m_k m_l +
\sum_{ik}\epsilon_{ik}\left\{ \frac{(n^*+1)^2}{4} 
\left[ \left( 1 - m_i \right)
\left( 1 - m_k \right) + 1 \right] 
+ \frac{2}{\beta U} \right\}
\end{equation}
The local order parameter $m_i$ can be evaluated by the extremum condition for
free energy $\frac{\partial A}{\partial m_i}=0$. Taking into account the 
equation (\ref{lambda}) for $\beta J=0$
\begin{equation}
\label{l-m}
\lambda_i(0) =  \frac{\partial A_1}{\partial m_i} 
\end{equation}  
Since $m_i$ is fixed equal to $<\sigma_{1i}>$ for any reduced hopping 
amplitude $\beta J$, it is in particular equal to  $<\sigma_{1i}>$ when 
$\beta J=0$, which gives us 
the important relation
\begin{equation}
\label{m-l}
m_i=\frac{\sinh\left(\lambda_i(0)\right)}
{\cosh\left(\lambda_i(0)\right)+\gamma(\beta U)}
\end{equation}
From (\ref{l-m}) and (\ref{m-l}) and from the homogeneity hypothesis on the 
interaction metrix ( $\epsilon_{ij}=1$ or $0$ ) we obtain a self-consistent 
equation for the order parameter $m_i=m$. Taking into account in $A_1$ only 
the first order we obtain  
\begin{equation}
\label{orderpar}
m=\frac{\sinh\left(2JD\beta m (n^*+1) \right)}
{\cosh\left(2JD\beta m (n^*+1) \right)+\gamma(\beta U)} 
\end{equation} 
Where D is the number of next neighbours sites. In the strong coupling limit 
we can neglect $\gamma$ in (\ref{orderpar}) and when $n^*=0$ we 
obtain the classical result of the mean field approximation for the hard core 
bosons model. As we shall see in the following, the knowledge of the expansion
up to the second order in the reduced hopping amplitude $\beta J$ becomes 
important to determine the relaxation of local perturbation\cite{FEM}. 

As far as the behaviour of the LQCP is concerned we note that in the 
non superfluid phase only the statistical average of $\sigma_{i0}$ does not 
vanish and then LQCP becomes trivial $(\rho_i(\pi_i)=\frac{1}{2})$. On the 
other hand it is easily seen that in the superfluid phase the  spontaneous 
symmetry 
breaking implies a non trivial LQCP $\rho_i(\pi_i)=\frac{1}{2}\left( 1 + m_i 
\cos \theta_i \right)$. Following the work of reference \cite{FEM} it is 
convenient, for the implementation of a quantum computer on a superfluid phase,
to identify the 0-logical state in the i-th site with the dependence of 
$\rho_i$ on $ \cos \theta_i $. To obtain a 1-logical state we need a local 
perturbation of LQCP. We consider the effect of an istantaneous local 
variation of the chemical potential asssociated to a perturbation Hamiltonian 
$H_p(t)= \delta(t-0^+) H_p$ with $ H_p= \sum_i n_i \theta_i$.
The time evolution of LQCP can be obtained in terms of the time evolution 
of the statistical average of time dependent raising and lowering operators. 
\begin{equation}
\rho_i^{\pm}(t)=\frac{tr \left( e^{-\beta H} exp\left( \sum_k 
\lambda_k\sigma_{1k} \right) e^{iH_P} \sigma^{\pm}_i(t) e^{-iH_P} \right)}
{tr \left(e^{-\beta H}exp \left( \sum_k \lambda_k\sigma_{1k} \right)  \right)}
\end{equation}
Where $\sigma_i^{\pm}(t)$ are operators in the Heisenberg representation.
Simple results are obtained assuming that the impulsive 
perturbation acting on a given site has no influence on next neighbours sites,
and if the evolution can be approximated to the first order in the reduced 
hopping amplitude. 
\begin{equation}
\ln \rho_i^{\pm}(t) = \ln \rho_i^{\pm}(0^+) \pm i t h_i
\end{equation}
where $h_i=J \sqrt{n^*+1} \sum_k m_k \epsilon_{ik}$ has the meaning of the 
internal static 
field if m is the static order parameter. At this stage of approximation we
obtain a free ``rotation'' around the symmetry breaking direction completely
analogous to that found in the magnetic case. We expect that decoherence 
effects will appear when the correlation of different sites degrees of freadom 
are taken into account. An evaluation of decoherence can be obtained 
introducing, in analogy with the reference \cite{FEM}, a dynamical function
\begin{equation}
\label{functional}
G_i^{\pm}(\lambda,t)= \ln \left\{ Tr \left[e^{-\beta H} exp \left( 
\sum_k (\pm iJt\sqrt{n^*+1} \epsilon_{ik}+ \lambda_k) \sigma_{k1}\right) 
\right] \right\}
\end{equation}
We assume that $\rho_i^{\pm}(t)$ can be written as 
\begin{equation}
\rho_i^{\pm}(t)=\sin \theta_i \Lambda_i^{\pm}(t)
\end{equation}
The dependence on $\sin \theta_i$ can be considered as the manifestation of 
the presence of a 1-logical states in site i\cite{FEM}. The time behaviour of
1-logical states $\Lambda_i^{\pm}(t)$ are given in terms of the generating 
dynamical function.
\begin{equation}
\label{Lambda}
\Lambda_i^{\pm}(t)=  exp\left( G_i^{\pm}(t)-G_i^{\pm}(0)\right)
\frac{\partial G_i^{\pm}}{\partial \lambda_i} 
\end{equation}
The equation (\ref{Lambda}) is actually an extrapolation of the result 
obtained in the first order expansion in the reduced hopping amplitude 
$(\beta J)$. The calculation of the dynamical generating function can be done 
introducing a Legendre transform at fixed time and reduced interaction 
\begin{equation}
\label{legendre}
F_i^{\pm}(m,\beta,t) = G_i^{\pm}(\lambda,\beta,t)-\sum_k 
\lambda_k(\beta,t) m_k
\end{equation}
Following an analysis similar to that of reference \cite{FEM} we obtain 
\begin{equation}
F_i^{\pm}(m,\beta,t) \approx F_i(m,\beta) \pm iJt\sum_k \epsilon_{ik} m_k 
-\Gamma t^2 
\end{equation} 
where $\Gamma$ is equal to $J^2 (n^*+1)^2\sum_k \epsilon_{ik}(1-m^2) 
+ O(\beta U) $.
We see that within the present approximation decoherence decreases as the 
superfluid parameter saturates $(m^2 \rightarrow 1)$. 

In conclusion we have studied the superfluid phase of a boson system on a 
regular
lattice with a strong on site interaction. The degenerate groundstates of the
energy operator associated to each site are used to define a qubit. Hopping 
of particles among sites defines the interaction of the qubits with the 
physical environment. The superfluid phase is defined in terms of a generating
function associated to the local coherence operator. This operator defines 
the direction in the subspace of the degenerate groundstates along which 
symmetry breaking occurs. Non trivial properties of LQCP are found in the 
superfluid phase. Quantum gating on a single qubit is associated to the 
modification of LQCP due to a suitable local impulsive gauge perturbation.  
Decoherence appears only when correction to the mean field approximation are 
taken into account. Slowing down of decoherence is obtained as the superfluid 
order parameter approaches saturation. In other words this 
particular environment gives rise, in the superfluid phase, to an effective 
field along which qubits tend to align ( 0-logical states ). Excitations of 
qubits in the orthogonal direction ( 1-logical states ) will relax slower and 
slower as superfluid order parameter saturates.


\end{document}